\documentclass[sigconf]{acmart}
\AtBeginDocument{%
  }

\copyrightyear{2026}
\acmYear{2026}
\setcopyright{cc}
\setcctype{by}
\acmConference[UIST Adjunct '26]{The 39th Annual ACM Symposium on User Interface Software and Technology}{November 02--05, 2026}{Detroit, MI, USA}
\acmBooktitle{The 39th Annual ACM Symposium on User Interface Software and Technology (UIST Adjunct '26), November 02--05, 2026, Detroit, MI, USA}
\acmDOI{10.1145/3830397.3841842}
\acmISBN{979-8-4007-2855-6/2026/11}

\begin{document}

\title{\sysname{}: Externalizing Stable and Situational Preference as Manipulable Spatial Objects}


\author{DaEun Choi}
\orcid{0000-0002-4843-0486}
\affiliation{%
  \institution{KAIST}
  \city{Daejeon}
  \country{Republic of Korea}
}
\email{daeun.choi@kaist.ac.kr}

\author{Daniel Lee}
\orcid{0009-0005-3061-1827}
\affiliation{%
  \institution{Salesforce AI Research}
  \city{Palo Alto}
  \state{CA}
  \country{USA}
}
\email{d.lee2@salesforce.com}

\newcommand{\sysname}[0]{Tasteply}

\begin{abstract}
Most recommender systems treat user preferences as a single signal to infer and optimize, yet human preferences are a mix of stable preferences and situational preferences that turn on and off depending on context. Through \sysname{}, we externalize this preference structure as manipulable spatial objects. \sysname{} represents each situational preference as an attractor and the stable preference as an underlying global field. Users can adjust their feed by moving around the space and adjusting the strength of the global field, turning the preference structure that remains latent within the model into something they can see and act on directly to steer the recommendation algorithm. Through an exploratory study (N=5), we found this spatial control effective, and identified its limits and future direction as a means of manipulating preference-based recommendations.
\end{abstract}

\ccsdesc[500]{Human-centered computing~Interactive systems and tools}

\begin{teaserfigure}
 \includegraphics[width=\textwidth]{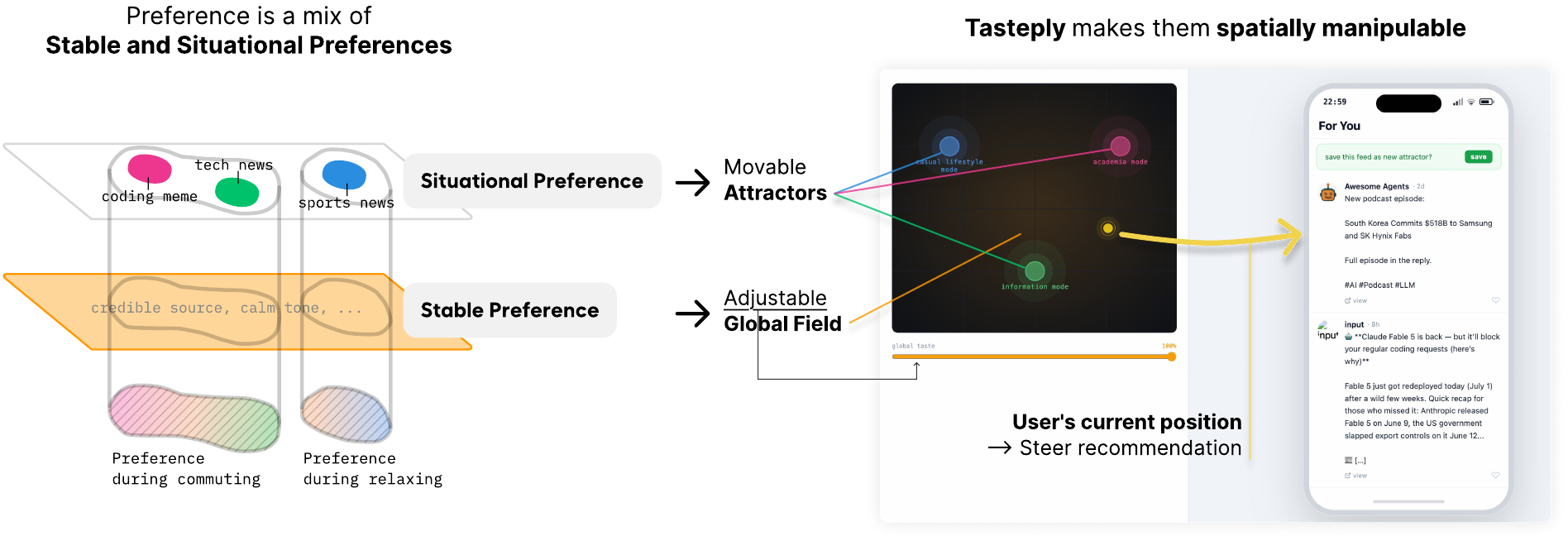}
  \caption{Human preference is a mix of stable preference that persists across contexts and situational preferences that activate by context (left). \sysname{} externalizes this structure spatially, representing each situational preference as an attractor and the stable preference as an adjustable global field, letting users steer their feed by moving among them (right).}
  \Description{Human preference is a mix of stable preference that persists across contexts and situational preferences that activate by context (left). \sysname{} externalizes this structure spatially, representing each situational preference as an attractor and the stable preference as an adjustable global field, letting users steer their feed by moving among them (right).}
  \label{fig:teaser}
\end{teaserfigure}


\maketitle

\section{Introduction \& Background}

People often get new information through algorithmic feed recommendations on platforms such as X, Threads, or Bluesky, which are built on a guess about what each user prefers.
Most recommender systems treat user preference as a single signal to infer and optimize~\cite{gru4rec2016, sasrec2018, bert4rec2019, peso2026}. Yet human preference is a mix of stable and situational components~\cite{li2005hybrid}: a stable, context-independent taste coexists with situational preferences that activate by context. For example, the same person may prefer light, humorous posts when browsing for fun, but dense, credible sources when researching for work.

Prior work captures this structure internally as multiple vectors~\cite{mind2019, comirec2020}, but never exposes it to users. Other systems let users adjust recommendations through sliders~\cite{tasteweights2012}, natural-language edits~\cite{radlinski2022, preferencediscerning2025}, or graphs~\cite{peerchooser}, but what users manipulate is a proxy for their preference, not a preference unit itself. How can we directly represent stable and situational preference as objects users can act on to adjust their recommendations?

We built \sysname{}, a system that externalizes this structure as manipulable objects on a 2D space. Each situational preference becomes an attractor and the stable preference an underlying global field, so users can directly adjust what has stayed latent inside the model. An exploratory study with 5 participants examined how people define, manipulate, and blend these preference units.
\section{\sysname{}}

\sysname{} is a Bluesky\footnote{https://bsky.app/}-based feed generation system that decomposes user preference into attractors and a global taste field. Each attractor occupies a point on a shared 2D space, with the global taste field as a separate layer beneath it. Users steer their feed by moving their position among the attractors, which blends nearby situational preferences, and by adjusting the strength of the global field. As they move, the feed updates in real time to reflect their position and chosen field strength. This spatial interaction follows a broader paradigm investigated in other domains~\cite{patchview, soo2005dust, ImaginationVellum}.

\subsection{Establishing the Preference Space}

\sysname{} builds a global taste field and an initial set of attractors from the user's liked posts, collected from Bluesky or exported from other platforms. Posts are clustered by content, and \sysname{} suggests an initial grouping of them into attractors that users can review and adjust. Each attractor is designed to be defined by a user's underlying motivations rather than surface-level topical similarity~\cite{zhao2026behavior}. As topic similarity alone cannot determine which context a user has in mind for a given post, the user's review and adjustment lets \sysname{} better capture the user's actual preference structure. After setting up the attractors, the global taste field is defined as a residual preference signals that were not covered in any of the situational preference attractors.

\subsection{Steering Through Preference Space}

With this preference structure in place, \sysname{} generates a personalized feed of Bluesky posts drawn from a continuously updated post pool. Users steer their feed using the preference space in the side panel. They can move the cursor on the preference map to blend nearby attractors. This gives users finer control over their feed than selecting from a fixed set of topics would allow. A global field slider adjusts how much the global preference pulls the feed toward the user's general taste.

\subsection{Technical Detail}
An attractor $a_i$ occupies a fixed position $p_i \in [0,1]^2$ with a representative vector $v_i$the normalized mean of the embeddings of posts assigned to it. $v_{global}$ captures the component of liked posts not explained by any attractor: each liked post embedding has its projection onto every $v_i$ subtracted, and $v_{global}$ is the normalized mean of these residuals.

Cursor position $c$ weights each attractor by an isotropic Gaussian kernel,
so that distant attractors contribute negligibly. The query vector $v_{query}$ is then calculated as the weighted sum of attractor vectors, renormalized to unit length.
Finally, the score for each post is calculated as
\(\mathrm{score}(post) = \cos(v_{query}, post) \times
\cos(v_{global}, post)^{\gamma}\)
where $\gamma$ (\texttt{global\_strength}) is user-adjustable.









\section{Evaluation \& Discussion}
We recruited 5 HCI researchers (4 female, 1 male; age M = 27.6, SD = 2.07) who had 2–7 years of research experience and regularly used social media platforms (Bluesky, X, Threads). This ensured participants could both engage naturally with feed-based recommendation systems and critically reflect on interaction design during the study.
Each participant provided their own platform usage history via screenshots, which served as the seed data for defining their preference space.
Sessions consisted of 15 minutes of free feed exploration with \sysname{}, and a 25-minute semi-structured interview about \sysname{}'s representation of preference structure and its interaction paradigm.

\paragraph{\textbf{Validating the Layered Preference Structure.}}
Participant behavior confirmed that \sysname{}'s layered structure mirrors how people already organize their preferences. Participants readily grouped topics they usually explored together into their own attractors (P3, P4, P5), and separately identified a stable preference that held regardless of context, such as preferred tone (P1). Participants also revised their structures iteratively. P3 split and re-grouped attractors multiple times, converging on boundaries that matched their actual usage purposes. This suggests the manipulable-attractor paradigm was usable, not just legible.

\paragraph{\textbf{Value of Continuous Spatial Control.}}
\sysname{} lets users blend multiple situational preferences continuously by positioning between attractors, rather than discretely by selecting a single one. The value of this continuous blending differed depending on what the user was trying to do. When exploring without a fixed target, users wanted to control not which exact content appeared but how much unfamiliar content entered the feed, which aligned well with the continuous control. For example, P5 wanted their niche interest to surface only occasionally within other casual feeds, and found positioning attractors let them set that degree of exposure directly.

When pursuing a specific goal, however, the same interaction offered limited value. P4 had a very clear goal of finding specific information on their feed, so they preferred to lock onto a single attractor, so blending offered little value for them. They were skeptical of the granularity of the control, saying they could barely perceive differences between nearby positions. We observed that no single interaction served both cases, so users should be able to collapse continuous control into discrete selection when their goal requires it.

\paragraph{\textbf{Toward a Context-Gated Preference Learning Loop.}}
Participants suggested \sysname{}'s separation of situational preferences could solve a problem they recognized in everyday feed recommendations. When a recommender relies on a single shared preference signal, a passing interest from one context can quietly reshape recommendations across every other context as well. P5 recalled how briefly engaging with a niche interest during casual browsing kept it resurfacing, unwanted, across unrelated sessions. P1 proposed a concrete improvement to the model: user feedback (such as likes or clicks) should update only the attractors active at that moment, weighted by their contribution to the current feed, describing this as a context-gated version of RLHF (Reinforcement Learning from Human Feedback). Therefore, we suggest extending this preference structure into a long-term recommendation algorithm so that each attractor remains a faithful, separate record of the situational preference it was built for.

\begin{acks}
We thank all of our study participants for their insightful feedback.
\end{acks}

\bibliographystyle{ACM-Reference-Format}
\bibliography{main}


\end{document}